\documentclass[a4paper]{article}
\usepackage[affil-it]{authblk}

\usepackage{graphicx}
\usepackage{graphics}
\usepackage{amsmath,amssymb}
\usepackage[usenames]{color}
\usepackage{subfigure}
\usepackage{hyperref}

\newcommand{\be}{\begin{equation}}
\newcommand{\ee}{\end{equation}}
\newcommand{\bea}{\begin{eqnarray}}
\newcommand{\eea}{\end{eqnarray}}
\newcommand{\ben}{\begin{enumerate}}
\newcommand{\een}{\end{enumerate}}
\newcommand{\bit}{\begin{itemize}}
\newcommand{\eit}{\end{itemize}}

\newcommand{\la}[1]{\label{#1}}

\definecolor{BrickRed}{cmyk}{0,0.89,0.94,0.28}
\definecolor{MidnightBlue}{cmyk}{0.98,0.13,0,0.43}
\definecolor{DarkGreen}{rgb}{0.100806,0.495968,0.209979}
\definecolor{orange}{rgb}{0.587167,0.354498,0.146197}

\begin{document}

\title{Electron transport through mesoscopic junctions revisited }

\author{Robert Alicki\thanks{E-mail: \texttt{robert.alicki@ug.edu.pl}}}

\affil{ICTQT, University of Gda\'nsk, 80-952 Gda\'nsk, Poland}

\date{\vspace{-5ex}}

\maketitle


\begin{abstract}
Theoretical foundations of electron transport in mesoscopic systems, based on Landauer theory, Master equations or Onsager linear thermodynamics, are revisited to show that the noniteracting electrons model  is insufficient to describe  neither passive  transport, nor  generation of electromotive force (active transport). It is argued that  2-body electrostatic interactions  creating double layers and surface charge distributions are crucial for the electron transport through a junction. Phenomenological modifications of the passive transport formulas  based on the carefull analysis of the fundamental notions of chemical, electrostatic, electrochemical, build-in potentials, band bending and bias voltage, are proposed. On the other hand active transport can be generated  by a  self-oscillating double layer (a pump ) driven by an external heat, light or chemical energy source.

\end{abstract} 

\section{Introduction}
\la{sec:intro}

The textbook theories of passive devices like diodes, transistors (including their ``single-molecule'' versions) and active ones transforming light, heat or chemical energy into electricity, like photovoltaic, thermoelectric and electrochemical cells (or their biological analogs) are based on the stationary picture of independent electrons moving ballistically or/and  randomly in a certain `` energy landscape''. These models are quite successful in explanation of the junctions' phenomenology, at least for  passive transport phenomena,  for a price of numerous confusions concerning fundamental notions  like chemical, electrochemical, and electrostatic potentials, bias voltage, open circuit voltage, built-in potential, band bending, Fermi level, etc.  The case of active systems, i.e. those producing work, is even more confusing as it is impossible to  generate electromotive force (emf), and hence current flowing in a closed circuit, by stationary potentials.  To solve the inconsistencies of steady-state models some additional mechanisms  of selective contacts or energy-filters, which remind the idea of Maxwell demon, were introduced.  To summarize these attempts one can quote \cite{SState}:``steady-state devices do this conversion [heat to work] without any macroscopic moving parts, through steady-state flows of microscopic particles such as electrons, photons, phonons, etc''.

This stationary picture of energy transducers has been challenged by the present author and coworkers in a series of papers \cite{Markovian} -\cite{dynamic}. However, the arguments remain unnoticed by the community,  probably due to the fact that those devices were constructed and applied very successfuly, often for more than 200 years, apparently without any urgent need for a consistent theoretical explanation. Hopefully, the situation  may  change due to the recent interest in the study of heat to work conversion  in mesoscopic systems using extremally precise control and measurement procedures. In contrast to macroscopic devices the theoretical description of mesoscopic ones seems to be much simpler and based on more fundamental evolution equations  which can be derived from the underlying  quantum models of many-body systems. In particular, three most popular formalisms are revisited: the  Landauer theory, Master equations approach, and Onsager relations.
\section{Models of electron junction}
\la{sec:junction}
 
To discuss the validity of commonly used models describing electronic junctions one can discuss only the simplest version of the setting. It consists of two electronic reservoirs (left and right) being at equilibrium states  at the temperatures $T_L , T_R$ and chemical potentials $\mu_L , \mu_R$ connected by a junction.  The junction is described either by a potential barrier (scattering model of point constact) or a single state which can be occupied by an electron (quantum dot) (Fig.1). The electrons are treated as an ideal gas of spinless fermions (i.e. polarized by a strong magnetic field) with electron-electron interactions included in the effective self-consistent potential. 

\begin{figure}
	\centering
		\includegraphics[width=0.70\textwidth, angle=00]{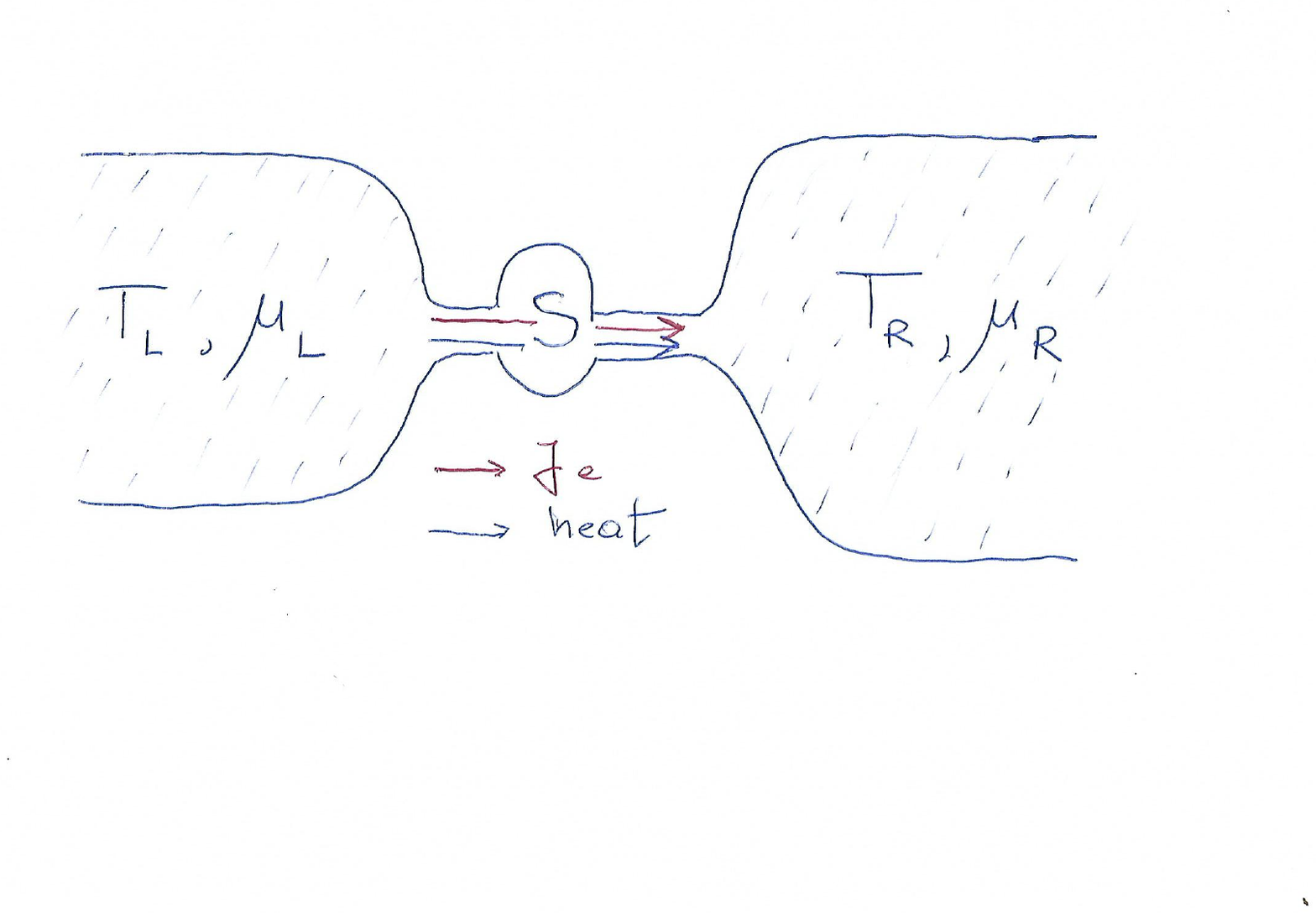}
	\label{fig:1}
\caption{\small Mesoscopic junction between two electronic reservoirs at the temperatures $T_L , T_R$ and chemical potentials  $\mu_L , \mu_R$ with $S$ denoting either point contact described by the scattering potential or a quantum dot with a single energy level. Models of noninteracting electrons with  $(T_L, \mu_L) \neq (T_R, \mu_R)$  predict nonzero electron current $J_e$. }
\end{figure}
%

\subsection{Landauer scattering model}

It is assumed that an electron travelling between two reservoirs is scattered by a potential at the junction and hence can be transmitted or reflected with the probabilities encoded in the ``scattering matrix'' $\mathcal{T}(E)$ which in this simple case is a complex-valued function of the electron energy $E$. Notice, that the scattering is fully reversible (unitary) and irreversibility of electric conduction  is hidden in the relaxation processes in reservoirs mantaining their equilibrium states. This is  similar to the Boltzmann equation approach where reversible two-body scattering process is combined with the product assumption for the N-particle probability distribution  (``Stosszahlanzatz'' or ``molecular chaos hypothesis''). Under these assumptions the standard derivation leads to the following expression for the the electron current (defined as number of electrons per unit time)  flowing from left to right reservoir
\begin{equation} 
J_e = \frac{1}{2\pi \hbar} \int_{-\infty}^{+\infty} dE\, |\mathcal{T}(E)|^2 \left[f_L (E) - f_R(E)\right] .
\label{landauer}
\end{equation}
Here
\begin{equation} 
f_X (E) = \frac{1}{ e^{(E - \mu_X)/k_B T_X} + 1}, \quad X = L , R 
\label{fermi}
\end{equation}
are Fermi-Dirac distributions for the reservoirs.
\par
Notice, that generically if the pairs of parameters $(T_L , \mu_L)$ and $(T_R , \mu_R)$  are different the model predicts nonzero current  flowing through the junction.


\subsection{Master equation approach}

For the quantum dot version of a junction model one can apply the quantum theory of open systems. The quantum dot can be treated as a 2-level system with the basis $|0\rangle , |1\rangle$ corresponding  to an empty and occupied dot's level, respectively. The dot is weakly coupled to both reservoirs by electron tunneling what allows to use the machinery of Markovian quantum open systems producing the Gorini-Kossakowski-Lindblad-Sudarshan Master equation for the reduced $2\times 2$ density matrix of the dot. The corresponding evolution transforms diagonal density matrices to the diagonal ones, and for our purpose of computing the stationary current it is sufficient to use rate equations for occupation probabilities $P_0 , P_1$ of the dot's states. They can be reduced to a single equation for $P_1(t)$
\begin{equation} 
\frac{dP_1}{dt} = -\left( \gamma_{\uparrow}+ \gamma_{\downarrow}\right) P_1 + \gamma_{\uparrow}
\label{master}
\end{equation}
where the excitation and decay rates are additive with respect to reservoirs
\begin{equation} 
 \gamma_{\uparrow}  =  \gamma^{L}_{\uparrow} +\gamma^{R}_{\uparrow} ,\quad  \gamma_{\downarrow}  =  \gamma^{L}_{\downarrow} +\gamma^{R}_{\downarrow} ,
\label{rates}
\end{equation}
and due to the equilibrium character of them satisfy the local detailed balance condition in the form
\begin{equation} 
\gamma^{X}_{\uparrow} = e^{-(E-\mu_X)/k_B T_X}\gamma^{X}_{\downarrow} ,  \quad X = L , R .
\label{kms}
\end{equation}
The values of decay rates can be calculated using Fermi Golden Rule for the tunneling interaction Hamiltonian with the averaged tunneling rates $g_X (E)$ such that, for example
\begin{equation} 
\gamma^{X}_{\downarrow} = \frac{2\pi}{\hbar} |g_X (E)|^2 \left[ 1 - f_X (E)\right],  \quad X = L , R .
\label{decayrate}
\end{equation}
The electron current from left to right at the stationary state  ($\bar{P}_1 =  \gamma_{\uparrow}/( \gamma_{\uparrow}+ \gamma_{\downarrow})$) is given by  
\begin{equation} 
J_e   =  - \left(\gamma^{L}_{\uparrow} +\gamma^{L}_{\downarrow}\right) \bar{P}_1 +  \gamma^{L}_{\uparrow} = \frac{  \gamma^{L}_{\uparrow} \gamma^{R}_{\downarrow} -\gamma^{L}_{\downarrow} \gamma^{R}_{\uparrow} }{\gamma_{\uparrow}+ \gamma_{\downarrow}}.
\label{ecurrent}
\end{equation}
Finally, inserting \eqref{kms} into \eqref{ecurrent} one obtains
\begin{equation} 
J_e   =  \frac{  \gamma^{L}_{\downarrow} \gamma^{R}_{\downarrow}}{\gamma_{\uparrow}+ \gamma_{\downarrow}}\left( e^{-(E-\mu_L)/k_B T_L} - e^{-(E-\mu_R)/k_B T_R}\right)  .
\label{ecurrent1}
\end{equation}
Again, like for the Landauer model, according to \eqref{ecurrent1} the electron current should flow through the junction except the nongeneric case when $(E-\mu_L)/ T_L=(E-\mu_R)/T_R$.


\subsection{Onsager theory}

If temperature and chemical potential differences,  $ \delta T = T_L - T_R$, $\delta\mu = \mu_L - \mu_R $ are small  one can express
the electron current in the lowest order approximation in the form of Onsager relation for the electric current
\begin{equation} 
J \equiv - e J_e  =  L_{ee}\delta\mu   +  L_{eh} \delta T, 
\label{onsager}
\end{equation}
where $L_{ee}$ describes Ohmic conductivity while  $L_{eh}$ accounts for the coupling between charge and heat transport. The corresponding formulas for those coefficients are easy to derive from \eqref{landauer} and \eqref{ecurrent1}. It is claimed that the Onsager relation \eqref{onsager} explains thermoelectric effect if $\delta\mu/e \equiv V$ is identified with the output voltage, in particular giving the expression for the open circuit (or stopping) voltage ($J = 0$)
\begin{equation} 
eV_{oc}= -\frac{L_{eh}}{L_{ee}} \delta T, 
\label{Voc}
\end{equation}
In the  Section 4 the results obtained from the simple models and discussed above are compared to the phenomenology of junctions. The first problem is the interpretation of model parameters $\mu_L, \mu_R$  which are usually called ``electrochemical potentials''  and their difference is interpreted  as  ``bias voltage'', $V = \mu_L - \mu_R$. The physics of various potentials entering theoretical description of junctions is studied in the next Section.
 

\section{The zoo of potentials}

Electrons in a solid state sample of arbitrary shape and composed of different materials can be described, in principle, by  the effective self-consistent Hartree-Fock Hamiltonian corresponding to a picture of ideal fermionic gas where each single electron is moving in the potential including averaged interaction with background positive ions and other electrons. Solving corresponding single-electron Schroedinger equation one obtains the set  $\{ \varphi_k(\mathbf{x}) , \epsilon_k \}$ of wave functions and eigen-energies labelled by the collection of quantum numbers $\{k\}$ which in the case of homogeneous materials consists of wave vector $\mathbf{k}$ from the first Brillouin zone, the value of electron $z$-spin  component and the label of  energy band.

Obviously, the eigenvalues $\{ \epsilon_k \}$  do not depend on the electron's position and hence the idea of ``band bending'' at the interface of different materials, frequently used in the literature \cite{BandBend}, is strictly speaking incorrect.
On the other hand single-electron wave functions  $\{ \varphi_k(\mathbf{x}) \}$ are position-dependent, reflecting inhomogeneities of the material such as those associated with the large variations of the  electron density around a junction.  Thus ``band bending" is at best a visual effect which can be seen in a plot of admissible electron energies, if the color intensity of the points along each straight, horizontal line representing a single-electron level  $\epsilon_k $ is proportional to $|\varphi_k(\mathbf{x})|^2 $  (see Fig. 2a).

\begin{figure}
	\centering
		\includegraphics[width=0.70\textwidth, angle=90]{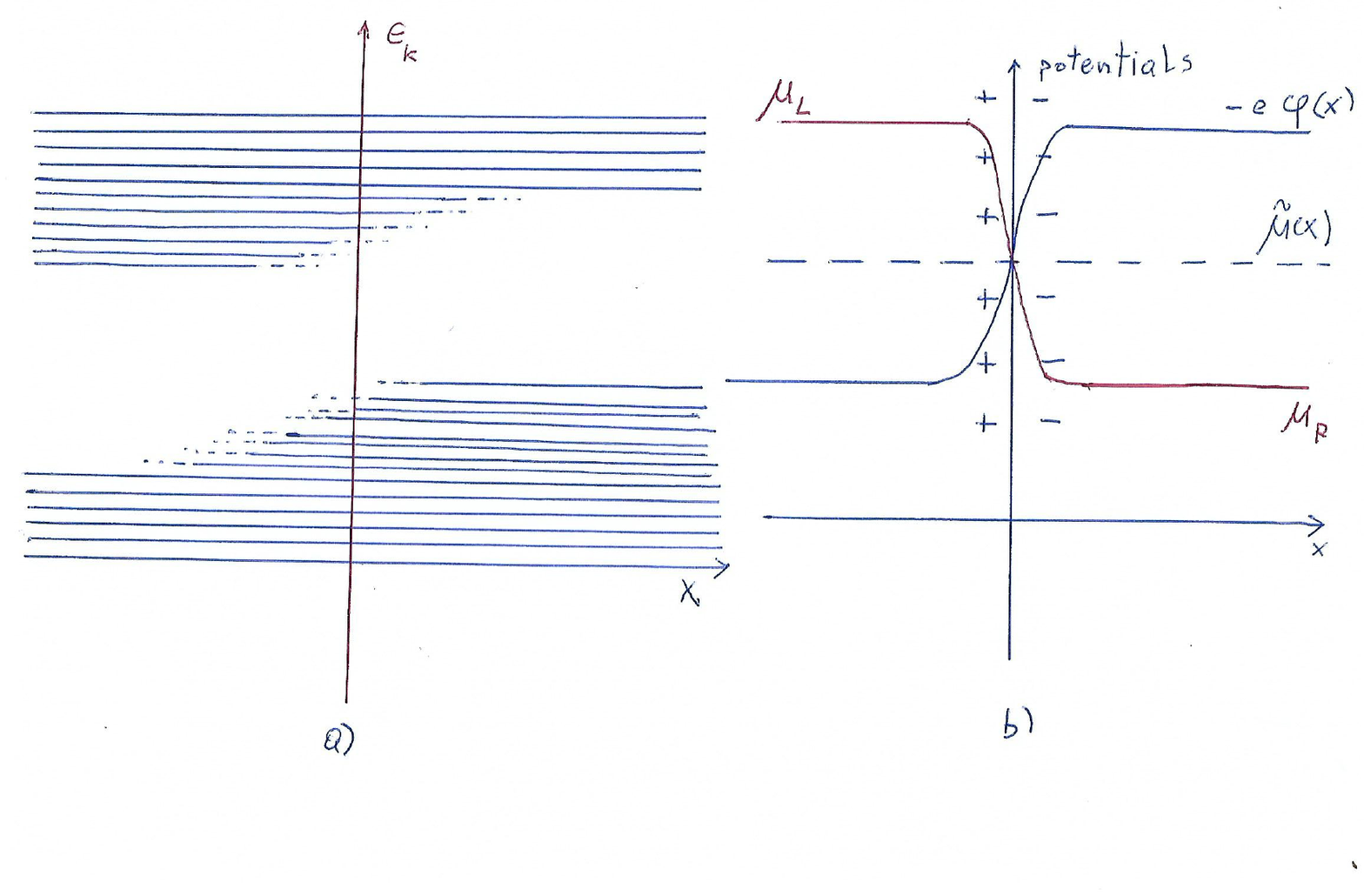}
	\label{fig:2}
\caption{\small ``Band bending `` and  potentials at the p-n junctions.\\
a) The fading horizontal lines reperesenting electronic energy levels $\epsilon_k$ illustrate the decreasing probability density for the corresponding wave functions $|\varphi_k({x})|^2 $ and create the visual effect  of ``band bending'' . They appear when a wave function corresponding to ``allowed energy'' in one material enters ``forbiden energy'' region in the other.\\
b) Creation of EDL. Difference of chemical potentials $\mu_L - \mu_R$ moves electrons to the right leaving uncompensated positive ions until the emerged build-in electrostatic potential $\phi(x)$ produces a constant electrochemical potential $\tilde{\mu}(x)$  of the equilibrium system.}
\end{figure}

For any sample of the volume $\mathcal{V}$ containing  $\mathcal{N}$  noninteracting electrons  at thermal equilibrium  the averaged occupation numbers  $n_k$ are given by the Fermi-Dirac distribution
\begin{equation}
n_k = \frac{1}{e^{\beta(\epsilon_k - \mu)/k_B T} + 1}
\label{F_D}
\end{equation}
with the temperature $ T$ and the global chemical potential $\mu =\mu(T)$ determined by the global density of electrons in the sample
\begin{equation}
\rho_e \equiv \frac{\mathcal{N}}{\mathcal{V}}=\frac{1} {\mathcal{V}}\sum_k \frac{1}{e^{(\epsilon_k - \mu)/k_BT} + 1}.
\label{densityg}
\end{equation}
It is crucial that even  for mesoscopic samples (like electronic reservoirs in discussed models)  $\mu (T)$ is a material constant ($\mu (0)$ is called \emph{Fermi level})  which does not depend on the excess charge distributed on the sample's suface which determines its constant  electrostatic potential  $\phi$ in the bulk. For a fixed  $\phi$  the total excess charge scales like a linear dimension of a sample giving a negligible contribution to electron density and hence to $\mu$, also.
\par
Position-dependence of electron wave functions $\varphi_k(\mathbf{x})$  allows to define  position-dependent  \emph{local density of electrons} $\rho_e (\mathbf{x})$ as well as position-dependent \emph{ local chemical potental} $\mu(\mathbf{x})$ 
\begin{equation} 
\rho_e (\mathbf{x}) =  \sum_k \frac{1}{e^{(\epsilon_k - \mu)/k_B T} + 1} |\varphi_k(\mathbf{x})|^2 \equiv
\frac{1}{\mathcal{V}}\sum_k \frac{1}{e^{(\epsilon_k - \mu(\mathbf{x}))/k_B T} + 1} .
\label{eldensity}
\end{equation}
In contrast to global $\rho_e , \mu$  local density $\rho_e (\mathbf{x})$ and local potential $\mu(\mathbf{x})$ contain  contributions from the surface charges generated by  electrons occupying surface states.
\par
The local electrostatic potential $\phi(\mathbf{x}) $ can be determined, in principle, by solving the Poisson equation with electron density \eqref{eldensity} and charge density of background ions. At electrochemical equilibrium, i.e. without external bias, we expect  that the \emph{local electrochemical potential} $\tilde{\mu}(\mathbf{x}) = \mu(\mathbf{x}) - e\phi(\mathbf{x})$,  ``smoothened'' over distances longer than the electron Fermi wavelength, is constant through the sample  (see Fig. 2b). This is consistent with the semiclasical  hydrodynamical picture of electron fluid driven by the gradient of electrochemical potential.  
\par
On the other hand,  the global electrochemical potential $\tilde\mu \equiv \mu -e\phi$   does not enter the Fermi-Dirac distribution because one  has to modify both: each eigenenergy $\epsilon_k \to \epsilon_k - e\phi$, and the chemical potential $\mu \to \tilde\mu = \mu - e\phi$.

\section{Junctions phenomenology vs. simple models}
\la{sec:phenomenology}

Two different materials, like metals (in normal or superconducting state) or semiconductors, both at the same temperature,  connected by a junction do not generate any stationary electric current despite the difference in their chemical potentials.  Of course, at the very begining some electrons move from the material where they are at a higher chemical potential (L) to the material in which they are at a lower potential (R) leaving uncompensated positive ions on the left. The created electrostatic double layer  (EDL) produces a built-in electrostatic potential which stops a further net electron transport  (see Fig.3a).  
\begin{figure}
	\centering
		\includegraphics[width=0.60\textwidth, angle=90]{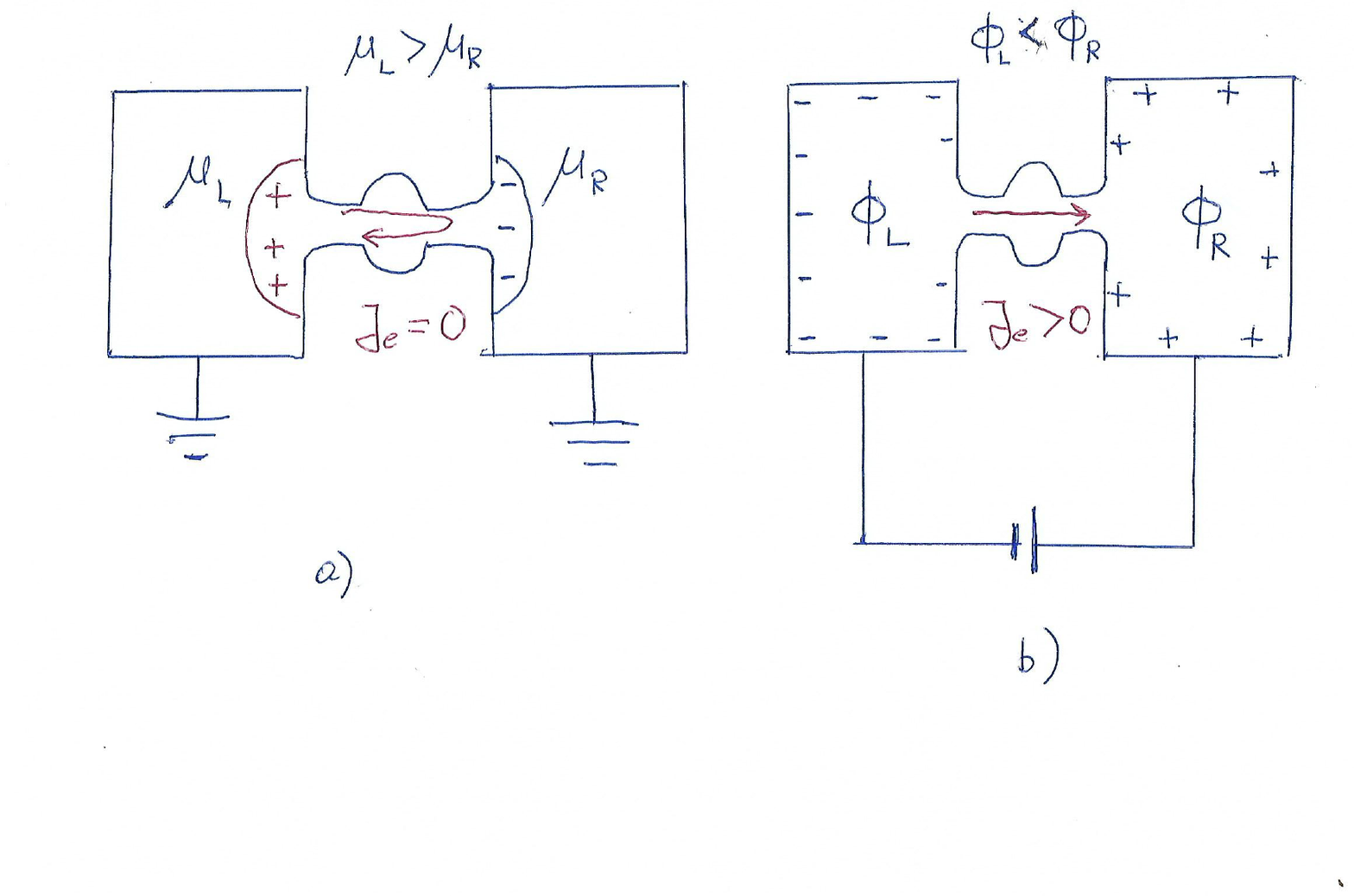}
	\label{fig:2}
\caption{\small Mesoscopic junction between two electronic reservoirs at the same temperature.\\
a) The case of electrically neutral system (grounded) with $\mu_L  > \mu_R$. Initially, electrons are transferred from $L$ to $R$  leaving behind uncompensated positive ions and forming  electrostatic double layer (EDL).  No steady current is flowing.\\
b) External bias generates surface charges and electrostatic bulk potentials $\phi_L , \phi_R$. A current proportional to  $V=(\phi_L - \phi_R)$ is driven.}
\end{figure}

\par
To drive a permanent current  one must apply an external bias to the system either attaching an external source of emf to reservoirs or charging them before connecting through the junction (Fig. 3b). The excess electric charge is distributed on the surfaces of  both materials producing constant electrostatic potentials $\phi_L$ and $\phi_R$ in the bulk. The generated current should be proportional to the bias voltage $V = (\phi_L - \phi_R)$ for small bias.
\par
It is clear that the simple models presented in the Section 2 do not give a consistent picture of the junction phenomenology. The main reason is that surface effects and surface electronic states responsible for generating built-in and bias potentials are not included in the models. The proper definitions and relations between chemical, electrochemical and electrostatic potentials were clarified in Section 3.
\par
When the temperatures of electronic reservoirs are different the electron current computed from the formulas \eqref{landauer},\eqref{ecurrent1} or \eqref{onsager} is interpreted, by the  proponents of ``steady-state conversion of heat into work'',  as a consequence of thermoelectric effect. However, the mechanism of creating built-in electrostatic potential which stops electron current through the junction is still valid.  One should stress that in the thermoelectric generator the electron current flows not only between two electronic reservoirs but also through the external load what means that emf  (work) is generated in the device (see Fig.4).   On the other hand, one can argue that work cannot be produced by  a steady electron current. Namely,  the power produced by the potential local force $\mathbf F (\mathbf{x},t)  = -\nabla \tilde{\mu} (\mathbf{x},t) $ acting on  electron current density $\mathbf {J} (\mathbf{x},t) $ , where $\tilde\mu (\mathbf{x},t) $ is an electrochemical potential, is given by \cite{LEC}
\begin{equation} 
P (t)=- \int d^3\mathbf{x}\,   \mathbf F (\mathbf{x},t) \cdot\mathbf  J (\mathbf{x},t)  = - \int d^3\mathbf{x}\,  \tilde\mu  (\mathbf{x},t) \frac{\partial \rho_e (\mathbf{x},t) }{\partial t}  .
\label{power}
\end{equation}
Here, $\rho_e(\mathbf{x},t) $ is the electron density which satisfies continuity condition
\begin{equation}  
\nabla\mathbf  J (\mathbf{x},t)  + \frac{\partial \rho_e (\mathbf{x},t) }{\partial t} = 0 . 
\label{continuity}
\end{equation}
It follows from \eqref{power} that steady power generation needs coordinated cyclic modulations of local electron density and electrochemical potential.  


\section{Modified equations for  passive transport }

In the previous sections it was  argued that Coulomb interaction between electron pairs and between electrons and background ions leads to creation of EDL at the  junction
which in the absence of external bias  completely stops  electron transport through any junction connecting two different materials at the same temperature $T$.  The EDL is a dynamical system itself with the equilibrium configuration (ground state) and oscillations around it (excited states).
\par
It is clear that the description of such complex system by first principles is difficult.  In the following a simple phenomenological construction is discussed, first for the case of passive transport. Because, the mechanism suppressing electric current due to EDL 
should not be influenced by the temperature difference between reservoirs one can discuss  passive transport in a more general setting, with two temperatures $T_L , T_R$.
\subsection{Passive transport in Landauer model}

The equation \eqref{landauer} is modified by using the different form of the Fermi-Dirac distributions
\begin{equation} 
J_e = \frac{1}{2\pi \hbar} \int_{-\infty}^{+\infty} dE\, |\mathcal{T}(E)|^2 \left[\tilde{f}_L (E) - \tilde{f}_R(E)\right] ,
\label{landauer1}
\end{equation}
where
\begin{equation} 
\tilde{f}_X (E) = \frac{1}{ e^{(E - \tilde{\mu}- e \phi_X)/k_B \tilde{T}} + 1}, \quad X = L , R .
\label{fermi1}
\end{equation}
Here, instead of  material constants $\mu_L , \mu_R$ we  use the single electrochemical potential $\tilde{\mu}$ which interpolates beween $\mu_L , \mu_R$ and gives a better approximation for the reservoirs electronic fluid parameters in the vicinity of the junction, modified by the presence of EDL.  Similarly, we can use a certain effective temperature $\tilde{T}$ interpolating between $T_L , T_R$. The new parameters are electrostatic potentials $\phi_L , \phi_R$ generated by the external bias (uncompensated surface charges), such that the current through the junction is proportional to the bias voltage $V =\phi_L - \phi_R$ in the linear regime. The precise values of   $\tilde{\mu}, \tilde{T}$  depend on the detailed physical properties of the sample and on the bias potentials $\phi_L , \phi_R$ also.


\subsection{Master equation for passive transport}

For the Master equation approach we can use the formula \eqref{ecurrent1} with  modified Boltzmann factors obtaining the following expression for electron current

\begin{equation} 
J_e   =  \frac{  \gamma^{L}_{\downarrow} \gamma^{R}_{\downarrow}}{\gamma_{\uparrow}+ \gamma_{\downarrow}}\left( e^{-(E-\tilde{\mu}-e\phi_L)/k_B \tilde{T}} - e^{-(E-\tilde{\mu}-e\phi_R)/k_B\tilde{T}}\right)  .
\label{currentmod}
\end{equation}
Here, the new parameters $\tilde{\mu} , \tilde{T} ,\phi_L , \phi_R$ possess the same meaning as in the modified Landauer formula \eqref{landauer1} and,  again, electron current is proportional to the bias voltage $V=\phi_L - \phi_R$ in the linear regime. 

\section{Thermoelectric effect as active transport }

If the temperatures of two external baths are different  one expects that the unbiased  junction may generate emf (and hence work) at the expense of heat absorbed from the hot bath. The geometry of the thermoelectric mesoscopic device is different from a simple model of the junction (Fig.1) as shown on Fig.4. One can use here a slightly modified scheme presented in the reference \cite{SState}, Fig.2b, where the heat source is coupled to  a quantum system interpreted now as a junction togather with EDL, while both electronic reservoirs are immersed in a cold bath.  The detailed analysis  of the  microscopic model has to be done, while here only a simplified phenomenological picture is discussed.

\begin{figure}
	\centering
		\includegraphics[width=1.3\textwidth, angle=00]{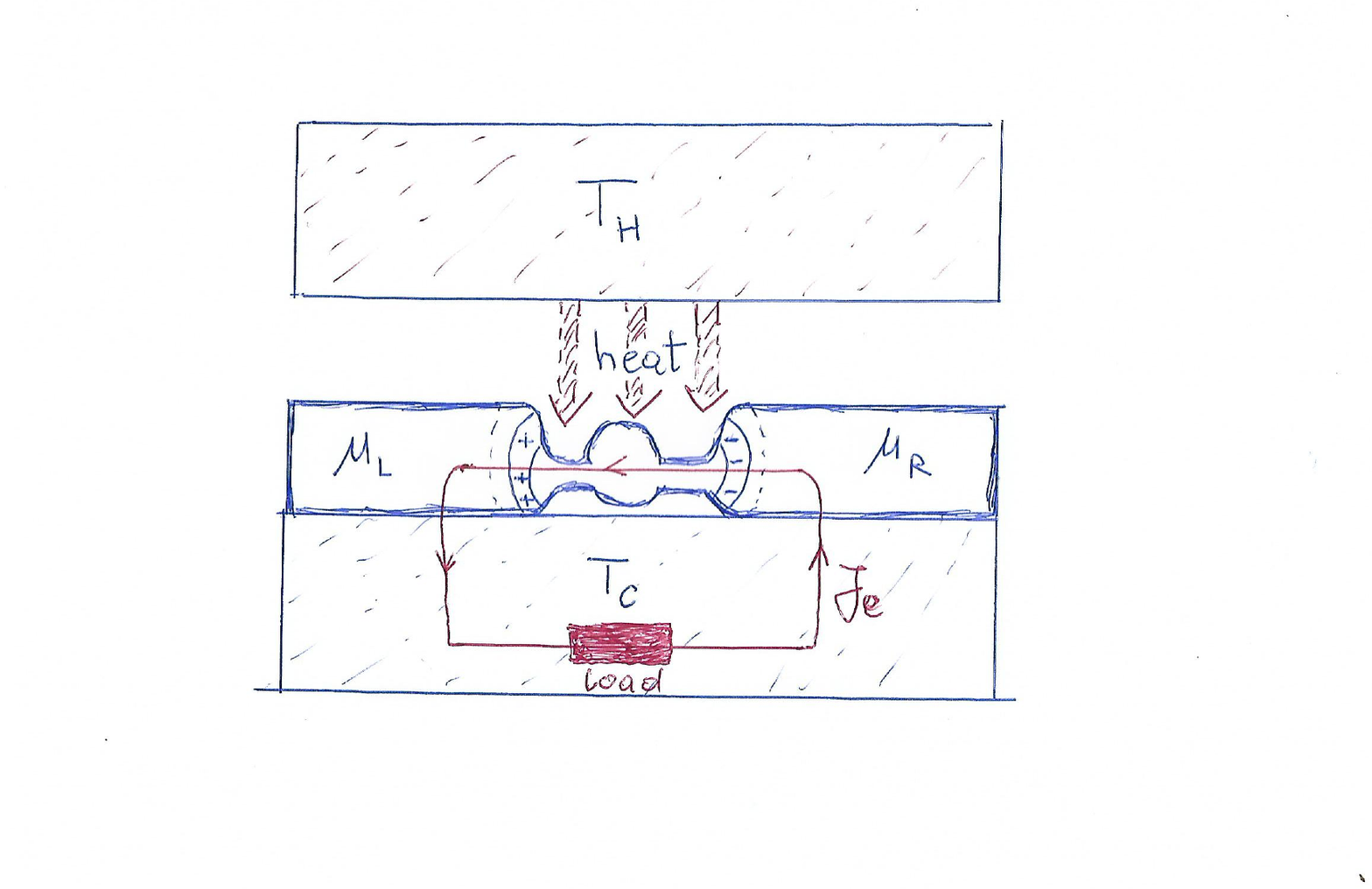}
	\label{fig:5}
\caption{\small EMF generation in a mesoscopic thermoelectric generator. Heat provided at the junctions drives selfoscillations of the EDL, which in turn pump an electric current in the closed circuit}
\end{figure}

EDL can be seen as  an open dynamical system coupled to two heat baths at  different temperatures: $T_H$ for a hot bath and $T_C$ for a cold one. By a suitable feedback mechanism, self-oscillations of EDL can be generated at the expense of heat provided by the hot bath, thus creating a heat engine part of the mesoscopic thermoelectric generator. The simple model of self-oscillations, which in  various versions has been used in \cite{Markovian},\cite{cycle},\cite{battery},\cite{dynamic},\cite{LEC}, consists of two equations
\begin{equation} 
 \ddot\xi + \gamma \dot\xi + \omega^2\xi =  \kappa ,
\label{newton}
\end{equation}
\begin{equation} 
\dot\kappa + \Gamma \kappa = -f \xi .
\label{kinetic}
\end{equation}
The equation \eqref{newton} is a linearized Newtonian equation for the electromechanical variable $\xi(t)$ representing deviation of EDL  from its equilibrium configuration and treated as a damped and driven harmonic oscillator with the angular frequency $\omega$. The damping rate $\gamma$ is due to the internal friction and a load, and the external force $\kappa(t)$ is a kind of ``pressure'' executed by the nearest environment.  Variations of $\kappa$ are caused by modulations of heat flow through EDL, which is driven by the temperature gradient $\Delta T = T_H - T_C$. In turn, heat flow modulations are generated by EDL oscillations. Such a feedback effect is acounted, in the lowest order approximation, by the RHS of \eqref{kinetic}, where $f$ is the \emph{feedback coefficient }, while $\Gamma$ is the relaxation rate for $\kappa$'s perturbations. The detailed analysis performed in \cite{LEC}, for a similar model, shows that if $f$ is larger than the critical one $f_c =\gamma(\omega^2 + \Gamma ^2 + \gamma\Gamma)$, then arbitrarily small initial oscillations of $\xi$ grow exponentially. In reality, they are stabilized by the nonlinear effects (not included in the linearized eqs.\eqref{newton}, \eqref{kinetic}) producing steady self-oscillations of EDL.
\par 
These self-oscillations perturb the equilibrium  charge distribution at EDL  and hence create an oscillating  weak ``internal bias potential'' $\delta\phi(t)\sim \xi(t)$ which generates time-dependent temporal electric current  expressed by nonlinear relations
\begin{equation} 
J_e  (t) =  L^{(1)}_{ee} \delta\phi(t) +  L^{(2)}_{ee} [ \delta\phi(t)]^2  + \cdots ,
\label{onsageractive}
\end{equation}
valid for both, Landauer and Master equation types of mesoscopic junction's models.
\par
Notice, that one has to go beyond the lowest order approximation, because the time average $\langle\delta\phi\rangle = 0$ and hence, the averaged actively pumped electronic current is given by the second order term
\begin{equation} 
 J_e  =  L^{(2)}_{ee} \langle[\delta\phi]^2 \rangle .
\label{currentactive}
\end{equation}
The energy of electro-mechanical EDL oscillations $\mathcal{E}(t)$ is proportional to $\langle[\xi]^2 \rangle \sim\langle[\delta\phi]^2 \rangle$ and its time variation is governed by the phenomenological equation
\begin{equation} 
\dot{\mathcal{E}} + 2\gamma \mathcal{E} = \eta J_h,
\label{Ekinetic}
\end{equation}
where $J_h$ is the heat current provided by the hot bath and transduced to energy of EDL's  oscillations with the efficiency $\eta < 1$.  Here, $2\gamma$ accounts for energy dissipation and the load associated with electric current pumping (compare with eq.\eqref{newton}). Therefore, the  energy  averaged over many cycles     $\bar{\mathcal{E}}$  is proportional to  $J_h$. On the other hand  heat current is proportional to the temperature gradient $\Delta T =T_H - T_C$. Thus combining all proportionality relations one obtains the final relation $J_e \sim \Delta T$ describing the thermoelectric effect.  This can be written in the following form (remember that electric current $J \equiv -e J_e$)
\begin{equation} 
 J  = - GS \,\Delta T,
\label{thermoelectric}
\end{equation}
with the standard parametrization of the proportionality coefficient, where $G$ is the electric conductance and $S$ is the Seebeck coefficient.  
\par
This model produces some quantitative predictions also. Firstly, because EDL's self-oscillations pump electrons against the local chemical potential gradient reducing the electrostatic built-in potential and the charge accumulated at  EDL (see Fig. 2b), the maximal output voltage (open circuit voltage $V_{oc}$), being the difference of local electrochemical potential at the contacts, cannot exceed the initial difference of global chemical potentials, i.e.
\begin{equation} 
V_{oc} \leq \mu_L - \mu_R .
\label{Voc}
\end{equation}
Secondly, the laws of thermodynamics applied to a heat engine provide the following limit (Carnot bound) on the output power for the thermoelectric generator
\begin{equation} 
P_{out} =  J\cdot V \leq  J_h \left(1- \frac{T_C}{T_H}\right) \simeq \frac{K }{T_H} (\Delta T )^2.
\label{powerout}
\end{equation}
In the formula \eqref{powerout} the Fourier law $ J_h \simeq K \Delta T $ is used,  where  $K$ is the thermal conductance of the junction.
\par
Summarizing, we arrived at the ``engine-pump model'' of mesoscopic thermoelectric generator which possesses ``(mesoscopic) moving parts'' similarly to the earlier models of macroscopic devices \cite{Markovian} - \cite{dynamic}. Notice, that reversing the thermodynamical cycle, by providing work through an external bias, we can use the same system as refrigerator (Peltier effect).

\section{Concluding remarks}
\la{sec:conclusions}
The transport of electrons in solids or plasma is very different from the neutral particle transport or heat conduction due to a large Coulombic energy created by any uncompensated local electric charge density. Therefore,  the ideal fermionic gas model applied to electrons in solids  can be misleading in particular for inhomogeneous systems like junctions or surfaces of conducting materials. The deceptively simple single-electron models of junctions like Landauer or Master equation approaches reveal, by a closer look, serious inconsistencies. There are covered by ambiguous interpretations of the model parameters, like for example, identification of  chemical potential difference with the bias voltage, despite the fact that (global) chemical potentials are material constants while  bias voltage is a controlled external parameter.
Another source of confusion is application of Onsager linear thermodynamics beyond its limits. The linear relations between thermodynamical forces and flows may correctly describe return to equilibrium for 
small initial deviations, while being insufficient in the presence of strong nonlinearities and feedback phenomena leading, for instance, to self-oscillations \cite{SO}. Treating neutral particle current, heat current and electric current on the same footing, despite their different physical nature, is also not justified.  In order to develop correct theories of passive and active charge transport we need more realistic, but unfortunately much more complicated models taking into account Coulombic interactions between charged particles and the related various surface effects. On the experimental side, one can expect that self-oscillations present in the active mesoscopic devices should be ``heard'' in electromagnetic or/and acoustic domain similarly to the observed emissions from mesoscopic Josephson junctions \cite{JJ1},\cite{JJ2} (see \cite{JJ} for a treatment of Josephson junction as an engine).
Last but not least, the author's  revisiting of  the theory of active transport has been motivated by the former observation that the very definition of work, for quantum open systems, demands  presence of  deterministic motion executed by an essentially classical degree of freedom  which can be modelled by time-dependent Hamiltonian \cite{Alicki79}.


\end{document}